\documentclass[aps,prb,twocolumn,groupedaddress,showpacs]{revtex4}
\usepackage{graphicx}

\begin{document}
\title{Shot noise of large charge quanta in superconductor/semiconductor/superconductor junctions}
\author{F. E. Camino}
\author{V. V. Kuznetsov}
\author{E. E. Mendez}
\affiliation{Department of Physics and Astronomy, Stony Brook
University, Stony Brook, New York 11794-3800, USA}
\author{Th. Sch\"{a}pers}
\author{V. A. Guzenko}
\author{H. Hardtdegen}
\affiliation{Institut f\"{u}r Schichten und Grenzfl\"{a}chen
(ISG-1), Forschungszentrum J\"{u}lich, D-52425 J\"{u}lich, Germany}

\date{\today}

\begin{abstract}
We have found experimentally that the noise of ballistic electron
transport in a superconductor/semiconductor/superconductor junction
is enhanced relative to the value given by the general relation,
$S_V=2eIR^2\coth(eV/2kT)$, for two voltage regions in which this
expression reduces to its thermal and shot noise limits. The noise
enhancement is explained by the presence of large charge quanta,
with effective charge $q^*=(1+2\Delta/eV)e$, that generate a noise
spectrum $S_V=2q^*IR^2$, as predicted in Phys. Rev. Lett.
\textbf{76}, 3814 (1996). These charge quanta result from multiple
Andreev reflections at each junction interface, which are also
responsible for the subharmonic gap structure observed in the
voltage dependence of the junction's conductance.
\end{abstract}

\pacs{74.40.+k, 74.45.+c, 74.50.+r}

\maketitle

In the last few years there has been an increasing interest in
measuring shot noise since it can give a more complete picture of
the physics involved in a system under study than that offered by
conductance measurements alone.\cite{Blanter} Hybrid
superconductor/semiconductor/superconductor (S/Sm/S) devices with
few one-dimensional channels, referred to in the literature as
superconducting quantum point contacts (SQPC), provide an example of
the potential usefulness of shot noise in that regard. It has been
predicted that in these devices the shot noise should be much larger
than the Poissonian noise $S_I=2eI$ generated by electrons of charge
\textit{e}, as if it were created by a large charge quantum of the
order of $2\Delta/V$ ($eV\ll\Delta$), where $\Delta$ is the energy
gap of the superconducting electrodes and \textit{V} is the applied
voltage across the device.\cite{Averin}

This large charge quantum can be seen as a consequence of a
phenomenon (Andreev reflection) occurring at the Sm/S interface when
the energy, \textit{E}, of a quasiparticle incident on the interface
from the semiconductor side, is inside the energy gap of the
superconducting electrode.\cite{Andreev} Under this condition, the
quasiparticle (e.g., an electron) cannot enter the superconductor
and cannot be reflected from the interface (assuming an interface
with zero potential height), hence, the only possibility is for the
quasiparticle to annihilate, with the production of a retroreflected
hole of energy \textit{-E} in the semiconductor side and a Cooper
pair on the superconductor side (exactly the inverse occurs if a
hole hits the interface).

In a SQPC with a bias \textit{V} across its electrodes, a quasiparticle coming from one of the electrodes
generates a chain of $2\Delta/eV$ $(eV<2\Delta)$ Andreev reflections, each pair of which transfers a charge $2e$
across the junction, until the last Andreev-reflected particle is injected into a quasiparticle level in the
opposite superconductor electrode. As a consequence, this chain process transfers a net large charge quantum
$q^*\approx(1+2\Delta/eV)e$, whose shot noise has been predicted by Averin and Imam to be $S_V=2 q^*IR^2$ [Ref.
2].

Shot-noise enhancement and an indication of large charge quanta have
been found experimentally in S/insulator/S tunnel
junctions\cite{Dieleman} and in S/normal metal/S
junctions.\cite{Hoss} Furthermore, quantitative confirmation of the
theory of shot noise in SQPC\cite{Naveh} was found in aluminum point
contacts,\cite{Cron} supporting the idea that multiple Andreev
reflections are responsible for dissipative charge transfer between
superconductors. There have also been reports of shot noise in
semiconductor-based junctions being enhanced. For instance, the shot
noise in a S/Insulator/Sm junction has been found to be twice the
Poissonian noise;\cite{Lefloch} and in a quasi-diffusive S/Sm/S
junction it has been shown that in the limit of incoherent multiple
Andreev reflections, the shot noise is enhanced due to an increase
in the electron temperature with respect to the lattice
temperature.\cite{Roche} However, to the best of our knowledge,
there has not been any evidence of large charge quanta in
semiconductor-based junctions, probably due to the very strict
demands required for that observation, namely, a large density of
high-mobility electrons and a high electron transparency of the Sm/S
interfaces.\cite{Blonder}

We report here that by paying special attention to materials and
device optimization we have been able to observe shot noise of large
charge quanta in S/Sm/S junctions. To this effect, we have used a
hybrid device that consisted of a two dimensional electron gas
(2-DEG) defined by modulation doping in an
In$_{0.53}$Ga$_{0.47}$As/In$_{0.77}$Ga$_{0.23}$As/InP
heterostructure.\cite{Hilde} The 2-DEG was bounded laterally by two
Nb contacts separated by a distance, \textit{L}, of 0.4 $\mu$m (see
Figs. 1a and 1b). The 2-DEG mobility and carrier density, measured
at 4.2 K, were 3.5$\times$10$^5$ cm$^2$/Vs and 6.6$\times$10$^{11}$
cm$^{-2}$, respectively. As a consequence, the electronic mean free
path, \textit{l}, and coherence length, $\xi$ (at 1.2 K), were 4.6
$\mu$m and 0.6 $\mu$m, respectively. Since \textit{l} and $\xi$ are
larger than \textit{L}, the electronic transport in our device is
ballistic and the probability of sustaining multiple Andreev
reflections (MARs) is high, provided that the interface has very
good transparency.\cite{Blonder} This condition was favored by
confining the 2-DEG within the In$_{0.77}$Ga$_{0.23}$As layer, which
itself presents zero Schottky barrier at the lateral
metal/semiconductor interfaces.\cite{Mead} In addition, the Nb
electrodes were deposited with an ion beam deposition system that
allowed in-situ cleaning of the semiconductor lateral wall prior to
the metal evaporation; this process has proven to be crucial for a
good transparency of the Nb/2-DEG interface.\cite{Thomas1}

\begin{figure}
\includegraphics[width=86mm]{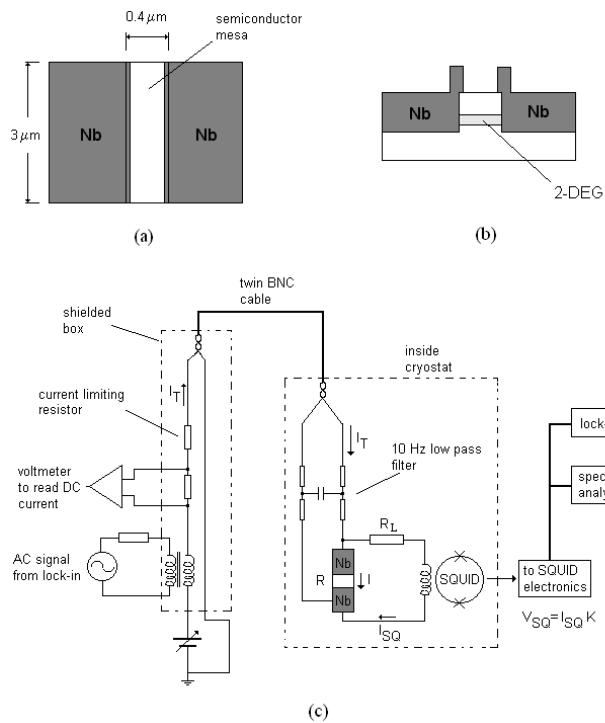}
\caption{\label{Schem} Schematic picture of the device and
measurement setup. (a) Top view of the hybrid device. The
semiconductor width, \textit{W}, and length, \textit{L}, are 3
$\mu$m and 0.4 $\mu$m, respectively. (b) Side view. The Nb
electrodes contact laterally the buried 2-DEG. (c) SQUID-based
measurement setup with an intrinsic noise of $\sim$ 0.5
pA/Hz$^{1/2}$. The junction resistance $R=dV/dI$ was measured
directly by injecting a small ac signal of 3 nA and $\sim$ 9 Hz
through the transformer and detecting the ac response across the
sample with a lock-in amplifier, after amplification by the SQUID
electronics.}
\end{figure}

\begin{figure}
\includegraphics[width=86mm]{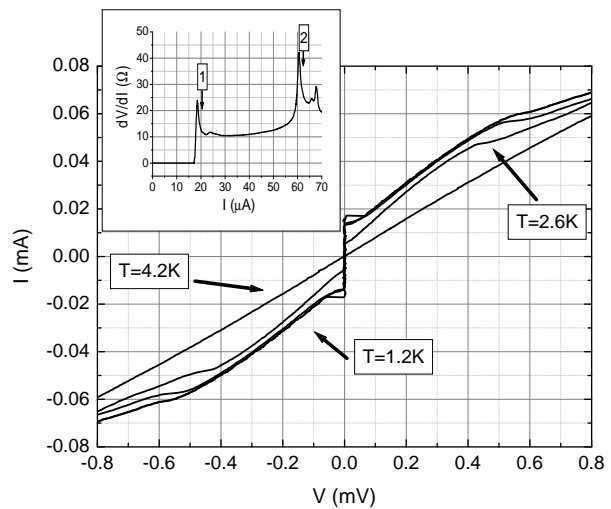}
\caption{\label{CVC} Current-voltage curves for the device sketched
in Fig. \ref{Schem} and described in detail in the text, measured at
several temperatures. Inset: $dV/dI$ vs. $V$ curve measured at 1.2 K
corresponding to a sweep down of the current. The arrows in the
inset point at the current (or voltage) regions selected for noise
measurements.}
\end{figure}

\begin{figure}
\includegraphics[width=86mm]{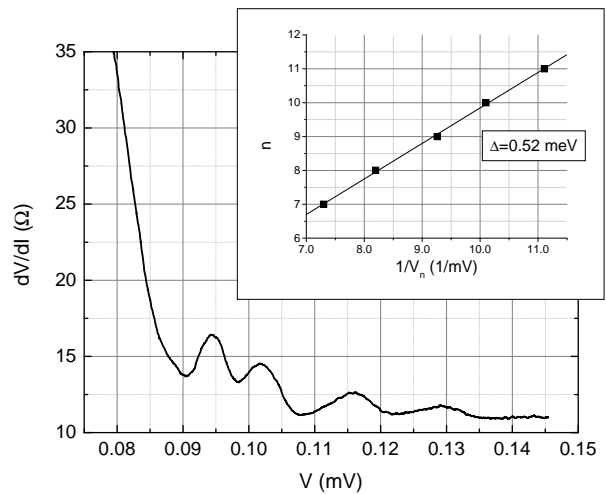}
\caption{\label{SGS} Dependence of resistance on voltage in region 1
of the inset in Fig. \ref{CVC}, measured at 1.2 K, for a downward
sweep of the current-voltage curve. We attribute the resistance
oscillations to subharmonic gap structure, from which we can
determine the energy gap $\Delta$, as shown in the inset. (See the
text for the method used to find $\Delta$.)}
\end{figure}

The measurement setup used, schematically depicted in Fig. 1c, was
based on a commercial (Quantum Design) superconducting quantum
interference device (SQUID). The SQUID proportionally transformed
the current circulating through its input coil into a voltage, with
a maximum gain, \textit{K}, of 4$\times$10$^6$ V/A (for $R_L=10$
$\Omega$). The current leads were of the twin BNC type, filtered at
the end close to the sample with an RC filter with cutoff frequency
close to 10 Hz. The filter, the sample (R), and the SQUID were
placed close to each other, shielded with a lead casing, and inside
a liquid-helium cryostat that could be pumped down to 1.2 K. To
reduce extrinsic noise during the measurements, the power supplies
for the voltage source circuit and DC current amplifier were battery
powered. For the same reason, the cryostat and measurement devices
were placed inside an RF-shielded room.

The results summarized in Figs. \ref{CVC} and \ref{SGS} provide
evidence that in our system Andreev reflection processes are
dominant, namely, the presence of excess current ($I_{exc}$) and of
subharmonic gap structure in the transport characteristics. The
current-voltage curves (CVC) in Fig. \ref{CVC} show a drastic change
with temperature, most notably from 4.2 K to 2.6 K. While at 4.2 K
the current is almost linear with voltage, as it corresponds to a
``normal'' metal (with resistance $R_N=13\:\Omega$), at 2.6 K and
below there appears a superconducting zero-resistance region at the
origin, followed by the onset of finite resistance when the current
exceeds the critical current $I_c\approx 17\:\mu$A. With increasing
current the resistance varies and even shows some structure, as
illustrated in the inset of Fig. \ref{CVC}. The current difference,
measured in the region of large voltages, between the CVC with a
superconducting state and the CVC with only a normal state is the
so-called excess current; its presence in the low-temperature
characteristics of Fig. \ref{CVC} is a clear indication of the
existence of Andreev reflections in our device.\cite{Blonder}

\begin{figure}
\includegraphics[width=86mm]{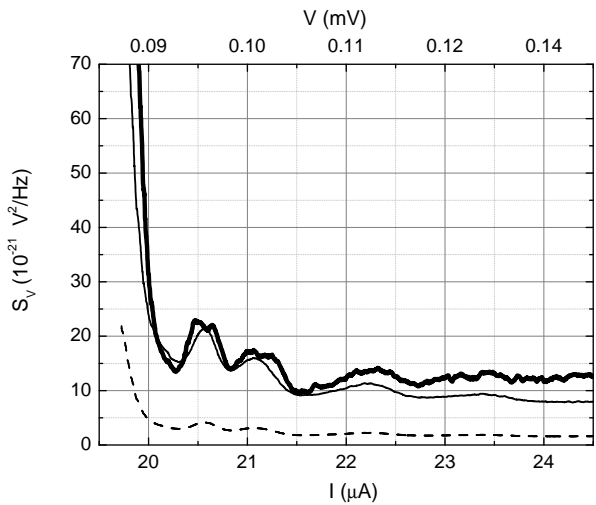}
\caption{\label{Noise1} Dependence of noise on voltage for region 1
(see inset in Fig. \ref{CVC}), measured at 1.2 K, corresponding to a
downward sweep of the current in the hysteretic current-voltage
curve. The thick solid line is the experimental curve, while the
dashed and thin solid lines correspond to plots using Eq.
\ref{eq:Classical} and Eq. \ref{eq:Quantum}, respectively. The value
of $\Delta$ = 0.52 meV used in Eq. \ref{eq:Quantum} was found from
the subharmonic gap structure (see text and Fig. \ref{SGS}).}
\end{figure}

\begin{figure}
\includegraphics[width=86mm]{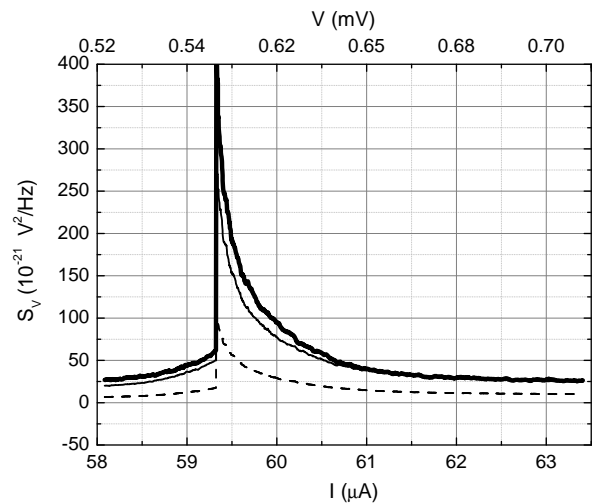}
\caption{\label{Noise2} Dependence of noise on voltage for region 2
(same considerations as for Fig. \ref{Noise1}). We observe that Eq.
\ref{eq:Quantum}, with the measured value of $\Delta$, reproduces
satisfactorily the measured curve without the need of any fitting
parameter.}
\end{figure}

The differential resistance, $dV/dI$, measured as a function of
voltage in the region near the superconducting-normal transition
(region 1 in the inset of Fig. \ref{CVC}) exhibits oscillations
known in the literature as subharmonic gap structure (SGS) and which
are consequence of multiple Andreev reflections.\cite{SGS1,SGS2} The
oscillations appear at voltages ${V_n = 2\Delta/ne}$, where $n$ is
an integer which corresponds to the number of Andreev reflections at
the Sm/S interface, and $\Delta$ is the superconducting energy gap
of the electrodes. By finding the slope in a plot of $n$ vs. $1/V_n$
(see inset of Fig. \ref{SGS}), we have obtained $\Delta$ = 0.52 meV.
The voltages of additional maxima in the device resistance, namely
\textit{V} = 0.6 mV (corresponding to the peak at \textit{I} = 60
$\mu$A in the inset of Fig. \ref{CVC}) and \textit{V} = 1.0 mV (not
shown), correspond, within experimental error, to the \textit{n} = 2
and \textit{n} = 1 elements, respectively, in the $V_n=2\Delta/ne$
series. Since each pair of Andreev reflections in a MAR chain
involves the transfer of a Cooper pair (of charge 2\textit{e})
across the junction, we can express the average transferred charge,
$q^*$, as a function of \textit{V} as $q^* \approx(1+2\Delta/eV)e$.

At this point, several remarks are in place. First, the value of
$\Delta$ determined from the SGS is considerably lower than the
value of $\sim$ 1.5 meV found in the literature for bulk Nb.
Although we do not have an explanation for this effect, we note that
reduced values of $\Delta$ have also been found in previous
works;\cite{Nguyen} in addition, the product $eI_cR_N$ = 0.22 meV is
comparable to the values we have found in similar junctions and
smaller than $\Delta$ as expected for this kind of
devices.\cite{IcRn} Second, the superconducting state of the device
occurred at a temperature ($<$ 4 K) well below the critical
temperature (7.5 K) of the Nb electrodes by themselves (measured
independently), which indicates the absence of electrical shorts in
the semiconductor region between the electrodes. This was confirmed
by inspecting the junction under a scanning electron microscope and
by performing an X-ray material analysis of the interelectrode
region.\cite{Camino} On the other hand, measurements of the CVC as a
function of magnetic field did not reveal the changes in the CVC
characteristics expected for electronic transport across the whole
width of our device,\cite{Thomas2} which indicates that in our
device the current flows through a small junction area. Our
structure approaches then the SQPC regime and it therefore seems
justified to interpret our experimental results in the light of the
theory described in Ref. 2.

The voltage noise measured as a function of current is shown in
Figs. \ref{Noise1} and \ref{Noise2}, for the two regions of current
labeled 1 and 2 in the inset of Fig. \ref{CVC} and measured at 1.2 K
and at a frequency of 3 kHz (see Ref. 20). Regions 1 and 2 in that
inset correspond to the thermal- and shot-noise limits,
respectively, of the well-established general relation for the
dependence of noise on \textit{V} and temperature \cite{Buttiker}
\begin{equation}
S_V=2eIR^2\coth(eV/2kT),  \label{eq:Classical}
\end{equation}
in which the cross-over from thermal ($S_V=4kTIR^2/V$) to shot noise
($S_V=2eIR^2$) occurs at around $eV=2kT$ (see Ref. 22). In both
regions, the measured noise (thick solid lines in Figs. \ref{Noise1}
and \ref{Noise2}) is significantly larger than that predicted
theoretically (dashed lines) for the two limits of Eq.
\ref{eq:Classical}, with enhancement factors of approximately 6 and
3, for regions 1 and 2, respectively.

Since our device unambiguously presents the signatures of multiple
Andreev reflections, as described above, we have interpreted the
enhanced noise as the shot noise of an effective charge, $q^*$,
along similar lines to those followed in S/insulator/S
junctions.\cite{Dieleman} In Eq. \ref{eq:Classical} we then replace
the electron charge, $e$, by the average transferred charge $q^*=
(1+2\Delta/eV)e$. After this substitution and using the value of
$\Delta$ = 0.52 meV mentioned above, the $\coth(q^*V/2kT)$ factor
becomes approximately one for the two current (or voltage) ranges
considered here. Consequently, the measured noise can be
approximated by the expression
\begin{equation}
S_V=2q^*IR^2,  \label{eq:Quantum}
\end{equation}
where the effective charge depends on voltage.

To test this analysis, we have plotted in Figs. \ref{Noise1} and
\ref{Noise2} (thin solid lines), the dependence of voltage noise on
current calculated using Eq. \ref{eq:Quantum}. As shown there, the
agreement with the measured values is very good throughout both
regions, and justifies our explanation of noise in terms of an
effective charge different from the electron charge.

The observation of enhanced shot noise in a S/Sm/S due to large
charge quanta opens the door to the study of shot noise in other
configurations in which Andreev reflections remain the main
mechanism for electronic transport. For instance, by adding a split
gate to the configuration studied here, it would be possible to
electrostatically tune in a continuous way the number of conduction
channels and test systematically the predictions for shot noise in
S/Normal/S junctions, from the single mode to the multimode
regime.\cite{Naveh, Cron} It would also be interesting to measure
the shot noise of S/Sm/S junctions with hot carriers injected
through separate electrodes. Since the supercurrent in a
multi-terminal S/Sm/S junction can be controlled by the injection of
hot carriers,\cite{Thomas3} it is reasonable to speculate that the
electronic noise might be affected as well, maybe reflecting a new
effective electronic temperature induced by the hot injection.

In conclusion, we have measured electron noise in a ballistic
superconductor/semiconductor/superconductor junction, and found it
to be enhanced with respect to the value given by the general
relation, $S_V=2eIR^2\coth(eV/2kT)$, for two voltage regions in
which this expression reverts to its thermal and shot noise limits.
Additionally, we have found that we can explain the measured noise
if we consider it as the shot noise, $S_V = 2 q^*IR^2$, of an
effective charge $q^*=(1+2\Delta/eV)e$, as predicted by
theory.\cite{Averin} These large charge quanta result from the
multiple Andreev reflection process which is responsible for the
subharmonic gap structure that we have observed in the $dV/dI$ vs.
$V$ curve, and from which we have determined the value of $\Delta$
used in the expression for $q^*$.

We would like to thank Profs. Richard Gambino and James Lukens from
Stony Brook University for letting us use their laboratory
facilities in the course of this work. This investigation has been
sponsored by the National Science Foundation, under Grant No.
DMR-0305384.

\end{document}